# Dynamical criticality of spin-shear coupling in van der Waals antiferromagnets

Faran Zhou[1], Kyle Hwangbo[2], Qi Zhang[1,2,9], Chong Wang[3], Lingnan Shen[2], Jiawei Zhang[1], Qianni Jiang[2], Alfred Zong[4], Yifan Su[5], Marc Zajac[1], Youngjun Ahn[1,6], Donald A. Walko[1], Richard D. Schaller[7], Jiun-Haw Chu[2], Nuh Gedik[5], Xiaodong Xu[2,3], Di Xiao[3,2], Haidan Wen[1,8]*

[1]X-ray Science Division, Argonne National Laboratory, Lemont, IL, USA
[2]Department of Physics, University of Washington, Seattle, WA, USA
[3]Department of Materials Science and Engineering, University of Washington, Seattle, WA, USA
[4]Department of Chemistry, University of California Berkeley, Berkeley, CA, USA
[5]Department of Physics, Massachusetts Institute of Technology, Cambridge, MA, USA
[6]Department of Materials Science and Engineering, University of Wisconsin-Madison, Madison, WI, USA
[7]Center for Nanoscale Materials, Argonne National Laboratory, Lemont, IL, USA
[8]Materials Science Division, Argonne National Laboratory, Lemont, IL, USA
[9]Present address: Department of Physics, Nanjing University, Nanjing, China.
*Correspondence to: wen@anl.gov

**Abstract**

The interplay between a multitude of electronic, spin, and lattice degrees of freedom underlies the complex phase diagrams of quantum materials. Layer stacking in van der Waals (vdW) heterostructures is responsible for exotic electronic and magnetic properties, which inspires stacking control of two-dimensional magnetism. Beyond the interplay between stacking order and interlayer magnetism, we discover a spin-shear coupling mechanism in which a subtle shear of the atomic layers can have a profound effect on the intralayer magnetic order in a family of vdW antiferromagnets. Using time-resolved x-ray diffraction and optical linear dichroism measurements, interlayer shear is identified as the primary structural degree of freedom that couples with magnetic order. The recovery times of both shear and magnetic order upon optical excitation diverge at the magnetic ordering temperature with the same critical exponent. The time-dependent Ginzburg-Landau theory shows that this concurrent critical slowing down arises from a linear coupling of the interlayer shear to the magnetic order, which is dictated by the broken mirror symmetry intrinsic to the monoclinic stacking. Our results highlight the importance of interlayer shear in ultrafast control of magnetic order via spin-mechanical coupling.

# Introduction

Understanding spin-lattice coupling is essential for deriving new control schemes of coupled degrees of freedom in quantum materials. In vdW materials, the unique spin-lattice coupling different from the three-dimensional system can lead to exotic electronic[1–4], magnetic[5–11], and mechanical[12,13] functionalities. Previous studies have shown that engineering structural configurations such as stacking order[14,15], interlayer twist angle[16], and strain states[17–20] can effectively control the interlayer magnetic coupling[15,21,22], thus tuning the interlayer magnetic orders. But the structural control of the intralayer magnetism can be challenging because the intralayer magnetic order is robust against small in-plane strain[23]. Enabling *interlayer* structural control of *intralayer* magnetism can expand the existing heterostructure engineering for controlling magnetism in all dimensions on demand, calling for a new spin-lattice coupling mechanism.

Here, we report that the shear motion of the atomic layers strongly couples to the intralayer magnetism, a phenomenon that is rooted in the unique structural anisotropy and magnetic symmetry of vdW materials. The interlayer shear, a degree of freedom that was largely overlooked due to the weak vdW interaction, turns out to play a critical role in determining the spin orders in individual layers. Our studies focus on a family of isostructural vdW antiferromagnets, $M$PS$_3$ ($M$: Fe, Ni, Mn)[24], which have monoclinic structures in the space group of C2/m with a monoclinic angle $\beta = 107.2°$. Below the Néel temperature ($T_N$), bulk FePS$_3$ and NiPS$_3$ exhibit the zigzag AFM order along the $a$ axis[25] while MnPS$_3$ exhibits the Néel-type AFM order (Fig. 1a). Although optical probes such as Raman spectroscopy[26–29] and second harmonic generation[30,31] have been applied to detect the local structural symmetry, a direct structural probe of long-range lattice configuration with atomistic accuracy is needed to understand the impact of crystallographic configurations on magnetic ordering quantitatively. Early equilibrium structural characterization has shown that multiple lattice degrees of freedom exhibit abrupt changes concurrently with the magnetic ordering temperature in FePS$_3$ [32]. However, it is not yet clear whether there is a primary structural parameter that couples with intralayer magnetism and how this lattice configuration influences the magnetism.

# Results

**Correlated critical slowing down of the interlayer shear and intralayer magnetic order**

To disentangle the complex coupling between multiple lattice degrees of freedom and magnetic orders in $M$PS$_3$, we drove the samples into nonequilibrium states by optical excitation and quantitatively tracked the structural and spin dynamics by x-ray diffraction (XRD) and transient optical linear dichroism (OLD), respectively (Fig. 1b). Time-resolved high-resolution XRD measurements were performed at the 7ID-C beamline of the Advanced Photon Source (see Methods). Three Bragg peaks $(002, \bar{2}02, 0\bar{2}2)$ of FePS$_3$ were measured as a function of time to reconstruct the dynamics of its unit cell. At the sample temperature of 95 K, below $T_N$ of 117 K, we discovered that the 002 peak shifted $\Delta q = 0.0098$ Å$^{-1}$ toward the -$\mathbf{c}$* direction at 5 ns (Supplementary Note 1 and Fig. 2b) as a result of an interlayer expansion upon optical excitation. This is consistent with an increase in lattice temperature by about 30 K (see Supplementary Note 2). Consequently, the interlayer expansion relaxed on the 200 ns time scale as the sample cooled down. This thermal recovery time increased as a function of temperature and peaked at 110 K. The magnitude and the peak temperature of this slowing down were reproduced using a thermal relaxation model involving the latent heat, as described in Supplementary Note 2.

Unlike the 002 peak that shifts toward the -**c**\* direction, the $\bar{2}02$ peak shifted to the opposite direction (+**c**\*) at sample temperatures below $T_N$. As shown in the schematic of the reciprocal space (Fig. 2a), the green and red parallelograms represent the unit cell in the *ac* plane before and after the laser excitation, respectively. The opposite shifts of 002 and $\bar{2}02$ peak correspond to a decrease in the monoclinic angle $\beta$ (Fig. 2a), i.e., there is an interlayer shear along the *a* axis after laser excitation. The dynamics measured at the $\bar{2}02$ peak (Fig. 2c) can be modeled by two simultaneous relaxation processes. One is the thermal relaxation that can be approximated by an exponential decay of the peak shift toward the -**c**\* direction. The other is the relaxation of the interlayer shear, which is proportional to the change of the monoclinic angle $\beta$ after the excitation as defined by $\Delta\beta(t) = \beta(t) - \beta(t<0)$ (Fig. 2d and Supplementary Note 3). The relaxation of $\Delta\beta$ can be approximated by an exponential decay of the peak shift to the +**c**\* direction. Therefore, a bi-exponential function fits the dynamics of the $\bar{2}02$ peak (solid lines in Fig. 2c and see Supplementary Note 4).

A striking observation is that the recovery time of $\beta$ is significantly longer than the thermal relaxation process measured by the 002 peak (Fig. 2e). The time constant of its recovery $\tau_\beta$ diverges at $116 \pm 2$ K, two orders of magnitude larger than the recovery time measured at 70 K. The error bar of the critical temperature represents uncertainties in determining the sample temperature under laser illumination. This critical temperature coincides with $T_N$ (117 K), indicating a strong correlation between the interlayer shear and the antiferromagnetic-to-paramagnetic phase transition. The relaxations of the interlayer shear along the *b* axis (angle $\alpha$, measured by $0\bar{2}2$ and 002 Bragg peaks) as well as the lattice parameters *a* and *b* (Supplementary Fig. 3) did not exhibit such critical slowing down. Therefore, the monoclinic angle $\beta$ is the only structural parameter of the unit cell that exhibits critical slowing down at $T_N$.

The divergence in the recovery time of the interlayer shear at the magnetic ordering temperature indicates a correlation of lattice shear with the spin order. To investigate the dynamics of the spin order, we performed time-resolved OLD measurements via transient absorption in a transmission geometry (Fig. 1b). The intralayer anisotropic optical absorption along and perpendicular to the zigzag spin chains results in a polarization rotation ($\varphi_{LD}$) of the linearly polarized optical probe beam. The change of polarization rotation ($\Delta\varphi_{LD}$) of the optical probe beam measures the dynamics of the intralayer zigzag AFM order[33,34]. To capture the recovery dynamics, we employed an asynchronous optical sampling technique (see Methods) which extended the measurement time window to microsecond scales and thus allowed us to accurately extract the relaxation time and the critical exponent. This time scale is otherwise challenging to access due to the limited time window of a few nanoseconds probed in an earlier study[33]. We found that the recovery time of the intralayer zigzag AFM order ($\tau_{LD}$) varies more than one order of magnitude as a function of temperature and diverges at $115 \pm 2$ K (Fig. 3a & b), indicating strong intralayer spin fluctuations at $T_N$.

The relaxation time of the interlayer shear ($\tau_\beta$) and the OLD spin reordering ($\tau_{LD}$) as a function of temperature were fitted by a power-law function $\tau = A * |1 - \frac{T}{T_c}|^{-zv}$ (Fig. 3b), where $A$ is the amplitude, $T_c$ is the critical temperature, $z$ is the dynamical critical exponent and $v$ is the critical exponent of the correlation length[35,36]. For comparison of the power-law scaling, the critical temperatures in each data set were adjusted to 117 K. The fitted *zv* values with fitting errors are 0.95±0.07 and 1.00±0.04 for interlayer shear and spin reordering, respectively, in good agreement with each other. We repeated the time-resolved XRD measurements in multiple samples with different thicknesses varying from 60 to 1300 nm (Supplementary Note 5). Although the absolute relaxation time scales are thickness dependent due to different thermal cooling time scales, all the power-law fittings yield the same critical exponent $zv \approx 1.0$ (Supplementary Table 1), suggesting that they belong to the same universality class. In addition to the agreement of the relaxation time, the amplitude change in the interlayer shear also correlates with the OLD change upon optical excitation as a function of temperature (Fig. 3c), further confirming their strong coupling.

Similar coupled critical slowing down was observed in the recovery of the intralayer AFM order and interlayer shear along the monoclinic stacking direction in NiPS$_3$, the same direction along the zigzag AFM spin chain (Fig. 3d). The power-law fit yields $zv \approx 0.33$, which is different from 1.0 in FePS$_3$ because NiPS$_3$ has an XXZ-type[9] rather than the Ising-type AFM order. Using a 2D XY model, the fit for the NiPS$_3$ data yields $v = 0.44 \pm 0.11$ (Supplementary Fig. 8), close to the theoretical value of $v = 0.5$ in the model[37,38]. In contrast, the same dynamical measurements of MnPS$_3$ revealed neither interlayer shear nor any critical slowing down across T$_N$ (Supplementary Fig. 7). The absence of critical slowing down in MnPS$_3$ suggests the important role of interlayer shear in the formation of intralayer zigzag AFM order because MnPS$_3$ does not host the zigzag AFM order but the Néel-type Heisenberg AFM order[24].

**Understanding the correlation between interlayer shear and antiferromagnetic orders**

Our multimodal ultrafast probes show that the interlayer monoclinic stacking, i.e., a subtle uniaxial translational displacement between layers, is highly correlated with the spin fluctuation. Close to T$_N$, the fluctuating intralayer AFM order prevents the recovery of interlayer shear, confirming the strong dynamical interplay of the interlayer shear and the intralayer zigzag AFM order.

To understand the coupled slowing down at T$_N$, we developed a phenomenological Ginzburg-Landau theory with two parameters: the magnetic order parameter denoted as $\eta$, which is related to the polarization rotation $\varphi_{LD}$ as $\eta^2 \propto \varphi_{LD}$ [34]; and the structural degree of freedom, denoted as $\delta\beta = \beta - \beta_{T>T_N}$, which represents the deviation of $\beta$ from its high-temperature value and is linearly related to the observable $-|\Delta\beta(t)|$ with an offset. An expansion of the free energy with respect to the two parameters around T$_N$ reads

$$F = A_\eta(T)\eta^2 + \frac{1}{2}B\eta^4 + \lambda\eta^2\delta\beta + A_\beta\delta\beta^2, \qquad (1)$$

where $A_\eta(T) = A(\frac{T}{T_N} - 1)$, $A$, $B$, $A_\beta$, and $\lambda$ are constants. As shown later numerically, the linear coupling between $\delta\beta$ and $\eta^2$ is essential in the critical slowing down of the shear motion. If the coupling is of the quadratic form $\eta^2\delta\beta^2$, no coupled critical slowing down is expected. For the interlayer shear along the *b* axis, the shear angle is defined as the change of angle $\alpha$: $\delta\alpha = \alpha - \alpha_{T>T_N}$. Because the mirror symmetry along the *b* axis requires $F(-\delta\alpha) = F(\delta\alpha)$, the linear coupling between $\delta\alpha$ and $\eta^2$ is forbidden. Therefore, no critical slowing down was observed in $\delta\alpha$.

The equilibrium solutions of $\eta$ and $\delta\beta$ can be obtained by minimizing the free energy $F$. The solutions take the form: $\eta = \pm\sqrt{2A_\eta A_\beta/(\lambda^2 - 2BA_\beta)}$ and $\delta\beta = -\lambda\eta^2/2A_\beta$ for temperature below $T_N$. Therefore, $\delta\beta$ is a higher-order small parameter with respect to $\eta$, and the dynamics around the Néel temperature is governed mostly by $\eta$. This correlation between $\delta\beta$ and $\eta^2$ in equilibrium is evident by comparing the measured $\delta\beta$ and $\varphi_{LD}$ as a function of temperature (Fig. 4a).

As the time scale of our measurement is much longer than the periods of magnon and shear modes, the time-domain response is in an over-damped regime without oscillatory dynamics. The numerical solutions (see Methods) of the time-dependent Ginzburg-Landau equation confirm that $\Delta\beta(t)$ follows $\Delta\varphi_{LD}(t)$ (Fig. 4b solid lines), while no coupled critical slowing down is observed for $\delta\alpha(t)$ since it couples to $\eta^2(t)$ quadratically (Fig. 4b red dashed lines). The scaling of the calculated relaxation time as a function of the temperature agrees with the experimental data (Fig. 4c).

The same symmetry principle also applies to $NiPS_3$. It exhibits a zigzag AFM order but the spins are mainly oriented within the layer (Fig. 1a), preserving the mirror symmetry with respect to the $a$ axis. Therefore, similar to $FePS_3$, a critical slowing down should be observed in the monoclinic tilt angle $\beta$ but not $\alpha$, which agrees with the experimental data (Supplementary Fig. 7). If the spin orientation deviates from the $a$ axis so that the mirror symmetry with respect to the $a$ axis is broken, a critical slowing down in $\alpha$ is also expected to occur. Our theory thus provides a guidance to understanding and engineering the linear coupling between magnetic orders and lattice symmetry, distinct from the quadratic coupling modeled in three-dimensional correlated materials[39,40].

## Discussions

The correlated slowing down of interlayer lattice shear and intralayer magnetic reordering clarifies their intimate relation that has been largely ignored. Although the structural change can be understood as the system accommodates the magnetic order by adjusting the lattice structures, it is interesting that these lattice degrees of freedom, $a$, $b$, $c$, $\alpha$, and $\beta$, are not equivalent in their relationship with the magnetism. $\beta$ shows a strong correlation with spin fluctuations because the linear term of $\eta^2\delta\beta$ survives in equation (1). What makes $\beta$ unique is that its linear coupling to $\eta^2$ is only allowed in the monoclinic structure. The interlayer shear of the monoclinic structure breaks the intralayer three-fold rotational symmetry and becomes the primary structural degree of freedom that couples to the zigzag AFM order with the same broken rotational symmetry (Fig.1a). As depicted in Fig. 4d, the coupled degrees of freedom jointly determine the potential energy surface. By driving the system out of equilibrium and following the relaxation pathways, the coupled parameters can be experimentally revealed. As the temperature increases close to $T_N$, the curvature of the energy surface is reduced, and the relaxation of coupled degrees of freedom slows down. Based on Fig. 4d, engineering the monoclinic angle away from its equilibrium value may systematically tune the free energy surface and thus influencing the magnetism. For example, decreasing monoclinic angle $\beta$ can reduce $\eta$. Inversely, one expects that the magnetic order becomes stronger by increasing $\beta$, which may be an effective control scheme to increase magnetic ordering temperature in vdW materials.

The dynamical measurements using pump-probe technique can directly interrogate nonequilibrium critical phenomena because the system is driven out of equilibrium and subsequently relaxes through a wide range of nonequilibrium states before returning to its ground state. The critical exponents as obtained by fitting the scaling of the relaxation time as a function of temperature thus reflect the intrinsic fluctuations of the non-equilibrium states. The critical exponents $z\nu$ obtained through fitting the power-law scaling in FePS$_3$ and NiPS$_3$ yield distinct values of $z\nu$, 1.0 and 0.3, respectively, reflecting their intrinsic nonequilibrium magnetic fluctuations. In FePS$_3$, the fitting result of $z\nu = 1$ is different from the equilibrium neutron scattering measurements that fits a 2D Ising model ($z\nu = 2.1$)[41]. This discrepancy suggests that the driven fluctuations in the critical region can be different from equilibrium fluctuations but may share similarities of driven system in 3D oxides[42].

In summary, through quantitative measurements of critical dynamics of structural and magnetic degrees of freedom at the Néel temperature of a family of vdW antiferromagnets, we uncover the interlayer shear as the primary structural degree of freedom that couples with the intralayer magnetic order in the vdW antiferromagnets. The phenomenological model provides a framework to describe linearly coupled degrees of freedom and guiding principles for designing correlated materials and dynamical processes with large spin-mechanical effects. Our work opens a new avenue for ultrafast structural control of magnetism in all dimensions, and inversely, for spin-mediated ultrafast mechanical processes.

## Methods

**Sample preparation and characterization.** Single-crystal $M$PS$_3$ bulk samples were mechanically exfoliated and transferred to sapphire substrates for XRD measurements. The OLD measurements were performed with samples on both the sapphire substrate and silicon nitride windows. The OLD results are similar with ~60 nm-thick sample on both substrates. The OLD data presented in this work are from samples on sapphire substrate. The typical lateral size of the flakes is on the order of 100 µm. Sample thickness varies depending on individual experiments. The thickness was measured by atomic force microscopy and the thickness fringes of high-resolution XRD radial scans.

**Time-resolved XRD experiments:** The time-resolved XRD experiments were performed at the 7ID-C beamline at the Advanced Photon Source (APS). The 400 nm pump laser pulse was derived by doubling the fundamental wavelength of an amplified Ti: Sapphire laser that outputs 100 fs laser pulse at 1 kHz repetition rate. The pump beam was incident normal to the sample with a spot size of 450 µm × 800 µm and a fluence up to 10 mJ·cm$^{-2}$. The x-ray pulse with a duration of 100 ps and 6.5 MHz repetition rate from the synchrotron was monochromatized to an energy of 11 keV with 0.1% bandwidth and focused to a size of 20 µm (vertical) × 60 µm (horizontal) by a pair of Kirkpatrick-Baez mirrors. The sample was mounted on a six-circle diffractometer (Huber GmbH.) and cooled down by a closed-cycle helium cryostat (Advanced Research Systems). An area detector (Pilatus 100K, DECTRIS Ltd.) was gated to selectively record the diffraction pattern from the x-ray pulses that pair with the laser excitation. The samples with various thicknesses from 60 to 1300 nm were measured. The pump fluence was adjusted so that the same thermal expansion ($\Delta c^*/c^* \approx 0.05\%$) was observed in samples with different thicknesses, which ensures the same optical-excitation effect regardless of the sample thicknesses.

**Ultrafast optical linear dichroism (OLD) experiments:** The ultrafast OLD experiments were performed at the Center for Nanoscale Materials (CNM). Asynchronous optical sampling technique was used to measure time delays up to microseconds[42]. The 400 nm, linearly polarized pump pulses were derived from a Ti: Sapphire laser with 80 fs in duration and delivered to the sample at 1 kHz repetition rate with a fluence of ~1.0 mJ·cm$^{-2}$. The probe pulse was a white light (400 nm to 850 nm range) generated by directing 1064 nm, 80 ps pulses through a photonic crystal fiber. The 1064 nm laser pulse is the output from a laser at a repetition rate of 2+Δ kHz synchronized with the pump laser, where Δ is the detuning frequency. The probe beam was linearly polarized and oriented about 45° relative to the sample *a/b* axis. The probe beam was transmitted through the sample and was split by a polarizing beam splitter into horizontal and vertical polarized beams, which were independently measured by a spectrometer with CMOS array detectors. The horizontal and vertical signals were analyzed to derive the polarization rotation angle $\Delta\varphi_{LD}$. The dynamics shown in Fig. 3a correspond to the spectrum response at 660 nm wavelength where the OLD effect reaches the maximum. A thin sample with a thickness of ~60 nm was used to allow sufficient optical transmission for detection.

**Numerical calculations based on the Ginzburg-Landau theory:** The time-dependent solutions of the Ginzburg-Landau equation are: $\partial\eta/\partial t = -\Gamma_\eta \partial F/\partial \eta = -2\Gamma_\eta(A_\eta\eta + B\eta^3 + \lambda\eta\delta\beta)$ and $\partial\delta\beta/\partial t = -\Gamma_\beta \partial F/\partial \delta\beta = -\Gamma_\beta(\lambda\eta^2 + 2A_\beta\delta\beta)$, where $\Gamma_\eta$ and $\Gamma_\beta$ are the damping rates. We first linearized the equation for $\partial\eta/\partial t$ by ignoring both the $B\eta^3$ and $\lambda\eta\delta\beta$ term. Approaching the critical temperature, $A_\eta(T) \to 0$ so that $\frac{\partial\eta}{\partial t} \to 0$. Thus $\eta^2(t)$ becomes time-invariant and exhibits a critical slowing down. Solving the equation for $\frac{\partial\delta\beta}{\partial t}$, we found $\delta\beta(t) = -\lambda\eta^2/2A_\beta + Ce^{-2A_\beta\Gamma_\beta t}$ with $C$ being a constant. Since the second term reduces to zero quickly due to large $\Gamma_\beta$, $\delta\beta(t)$ follows $\eta^2(t)$ on long time scales, i.e., $\delta\beta$ also exhibits a critical slowing down. The red and blue solid lines in Fig. 4b were calculated by numerically solving the time-dependent Ginzburg-Landau equation with the following parameters: $A_\eta = 0\ (T = T_N), B = 1.05, \lambda = 1.5, A_\beta = 1.3, \Gamma_\eta = 1.0, \Gamma_\beta = 1.3$. The red and blue dashed lines in Fig. 4b were calculated based on model: $F = A_\eta(T)\eta^2 + \frac{1}{2}B\eta^4 + \lambda\eta^2\delta\alpha^2 + A_\alpha\delta\alpha^2$, which has a quadratic order of δα in the coupling term thus results in no critical slowing down. The dashed lines were calculated based on the following parameters: $A_\eta = 0\ (T = T_N), B = 1.05, \lambda = 1.5, A_\alpha = 1.3, \Gamma_\eta = 1.0, \Gamma_\alpha = 1.3$. For the red solid line in Fig. 4c, the parameters are the same as above except $A_\eta$, which varies between -0.05 and 0.02.

**Data availability:** The data plotted in the main figures are provided in the Supplementary Information/Source Data file. Additional data are available from the corresponding authors upon request.

**Acknowledgments:** We thank Dr. John W. Freeland for discussion. This work was primarily supported by the U.S. Department of Energy, Office of Science, Basic Energy Sciences, Materials Sciences and Engineering Division, under Award No.DE-SC0012509 (Experimental design, data collection and analysis, theory, and partial manuscript preparation by F.Z., K.H., Q.Z., C.W., L.S., A.Z., Y.S., N.G., X.X., D.X., and H.W.). The use of facilities at the Center for Nanoscale Materials (R.D.) and Advanced Photon Source (D.W. and M.Z.), both U.S. Department of Energy Office of Science User Facilities, was supported by the U.S. DOE, Office of Basic Energy Sciences, under Contract No. DE-AC02-06CH11357. Bulk crystal growth (Q.J.) is supported by grant no. NSF MRSEC DMR-1719797 and the Gordon and Betty Moore Foundation's EPiQS Initiative, grant no. GBMF6759 to J.H.C. A.Z. acknowledges support from the Miller Institute for Basic Research in Science. H.W., J.W. and Y.A. also acknowledge the partial support for data collection and manuscript preparation by the U.S. Department of Energy, Office of Science, Basic Energy Sciences, Materials Sciences and Engineering Division.

**Author contributions:** F.Z., J.Z., M.Z., Y.A., D.W., A.Z., and Y.S. collected the ultrafast and static XRD data under the supervision by N.G. and H.W. F.Z., J.Z., K. H., Q.Z., and R.S. collected ultrafast optical linear dichroism data supervised by X.X and H.W. K. H. and F. Z. prepared the samples using the bulk compounds grown by Q. J. and J.H.C. C.W., H.S., and D.X. developed the theory. F.Z. and H.W. wrote the manuscript with input from all authors. H.W. conceived and supervised the project.

**Figures and Legends**

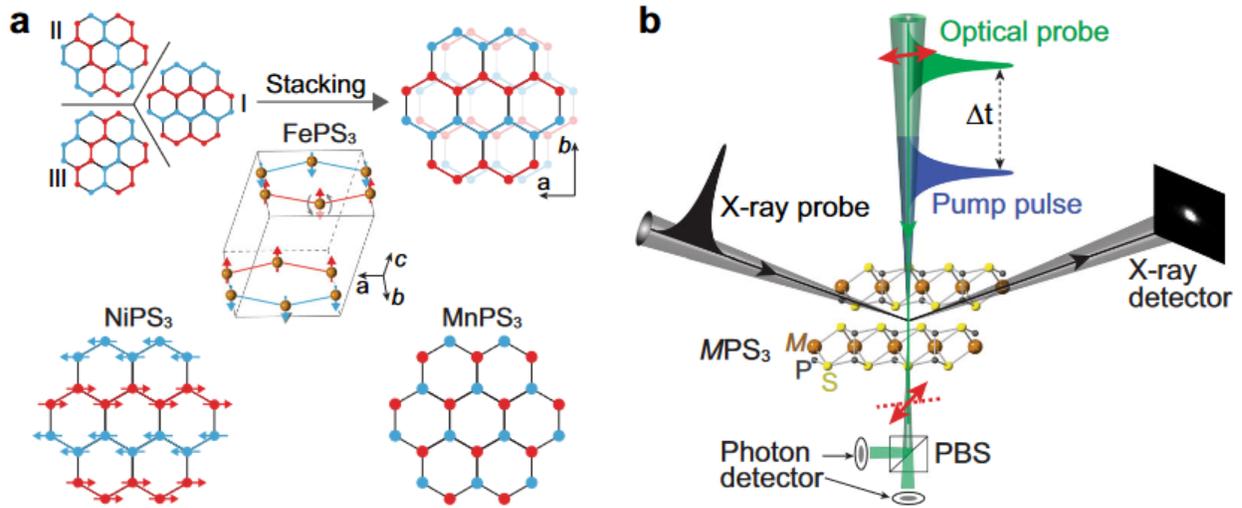

**Fig. 1 | vdW antiferromagnetic (AFM) materials and ultrafast measurements. a,** Top row: cartoons depict that the energy degeneracy of three intralayer zigzag antiferromagnetic orders is lifted by monoclinic stacking in FePS$_3$ so that the zigzag chain is stabilized along the monoclinic stacking direction. Red and blue dots represent up and down spin with respect to the basal plane. The cartoon in the middle row shows the three-dimensional unit cell of FePS$_3$, where only the Fe atoms are drawn. The curved gray arrows indicate the spin-flip fluctuations at random Fe atom sites close to $T_N$. Bottom row: magnetic structure of zigzag AFM in NiPS$_3$ and Néel-type AFM in MnPS$_3$. **b,** Time-resolved multimodal measurements of optically excited vdW antiferromagnets. The x-ray diffraction directly characterizes the lattice configuration while the optical transmission linear dichroism measures the intralayer AFM order. The red arrow indicates the polarization rotation of the optical probe light pulse. PBS: polarizing beam splitter.

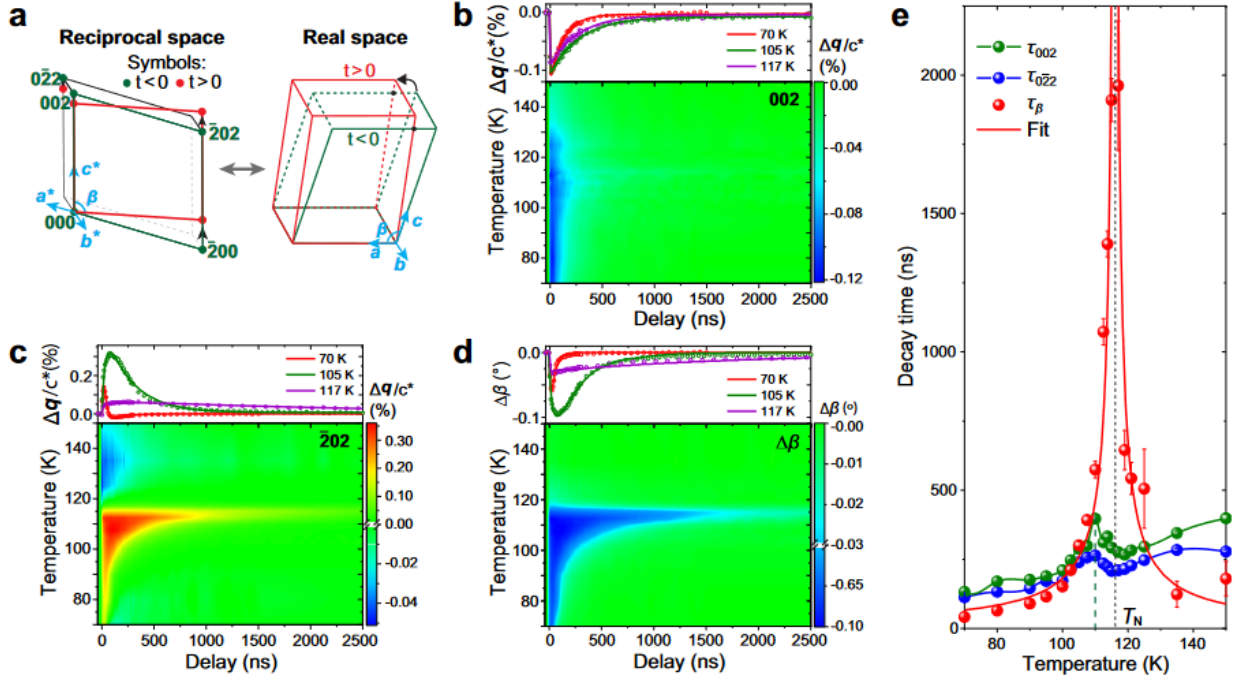

**Fig. 2 | Temperature-dependent structural dynamics in FePS₃. a**, Schematics of the reciprocal- and real-space unit cell before (green dots and lines) and after (red dots and lines) laser excitation. **b**, Dynamics of the 002 peak shift ($\Delta q$) along the **c**\* direction as a function of the temperature. **c**, Dynamics of the $\bar{2}02$ peak shift as a function of the temperature. **d**, Dynamics of $\beta$ change ($\Delta\beta$) as a function of the temperature. Top panels in (**b**, **c**, **d**) are the line cuts of three representative temperatures, in which the solid lines represent the fitting results (see text). **e**, Relaxation time of the 002, $0\bar{2}2$ peak shifts and $\Delta\beta$ across $T_N$. $\tau_\beta$ was fitted by a power-law function detailed in Supplementary Note 5. The vertical green dashed line at 110 K indicates the peak of the thermal relaxation time. The error bars represent the standard deviation of exponential fitting detailed in Supplementary Note 4.

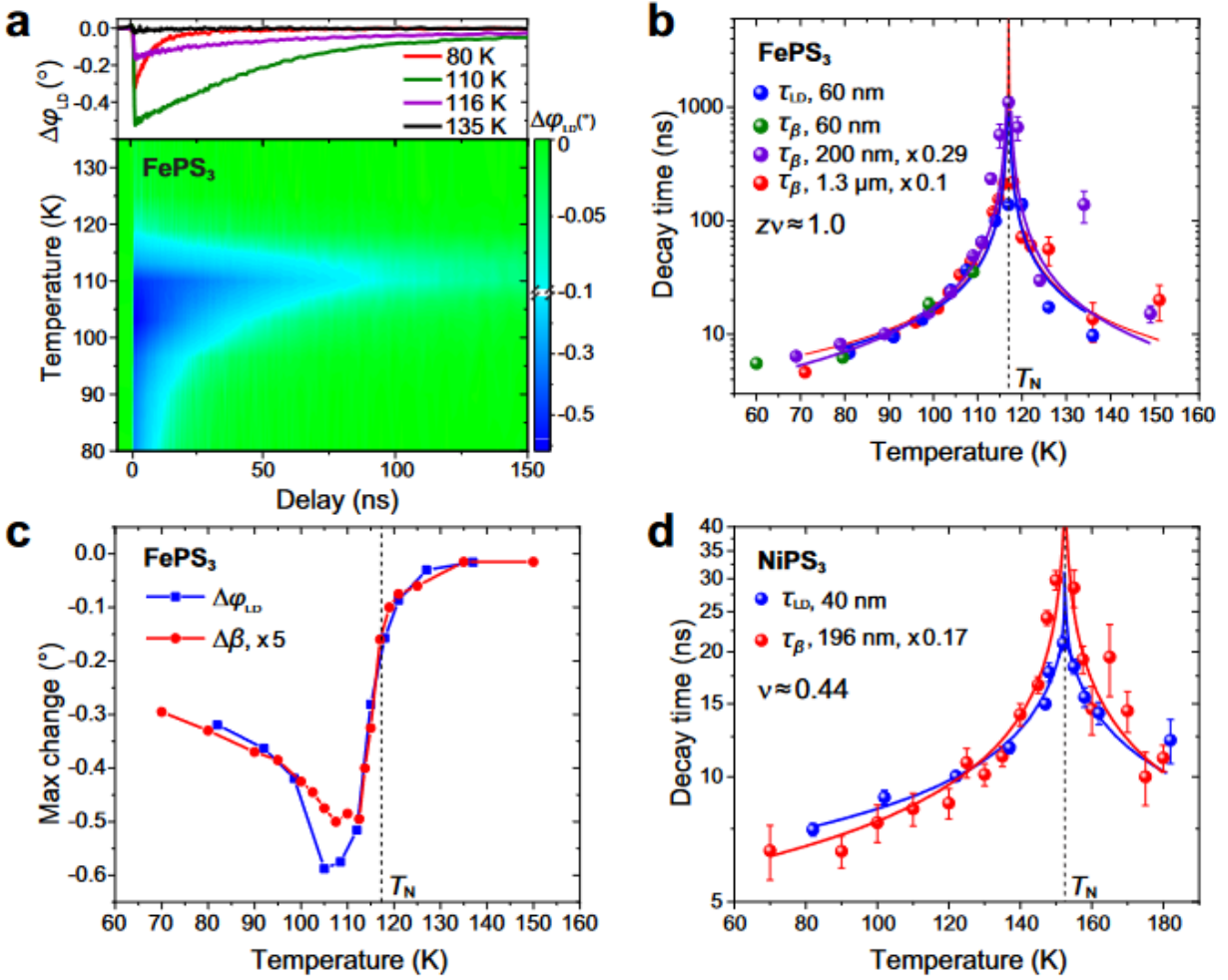

**Fig. 3 | Temperature-dependent spin dynamics in *M*PS$_3$. a**, Dynamics of OLD (polarization rotation change $\Delta\varphi_{LD}$). Top panel: line cuts at four representative temperatures of the color-map plot. **b**, Relaxation time of OLD and monoclinic angle $\beta$ in FePS$_3$. The sample thickness is labelled for each dataset. **c**, Comparison of $\Delta\varphi_{LD}$ and $\Delta\beta$ maximum change as a function of temperature. **d**, Relaxation time of $\Delta\varphi_{LD}$ and $\Delta\beta$ in NiPS$_3$ (See Supplementary Fig. 7 for raw data and Supplementary Note 5 for the fitting procedure). The error bars in **b** and **d** represent the standard deviation of exponential fitting results (see Supplementary Note 4).

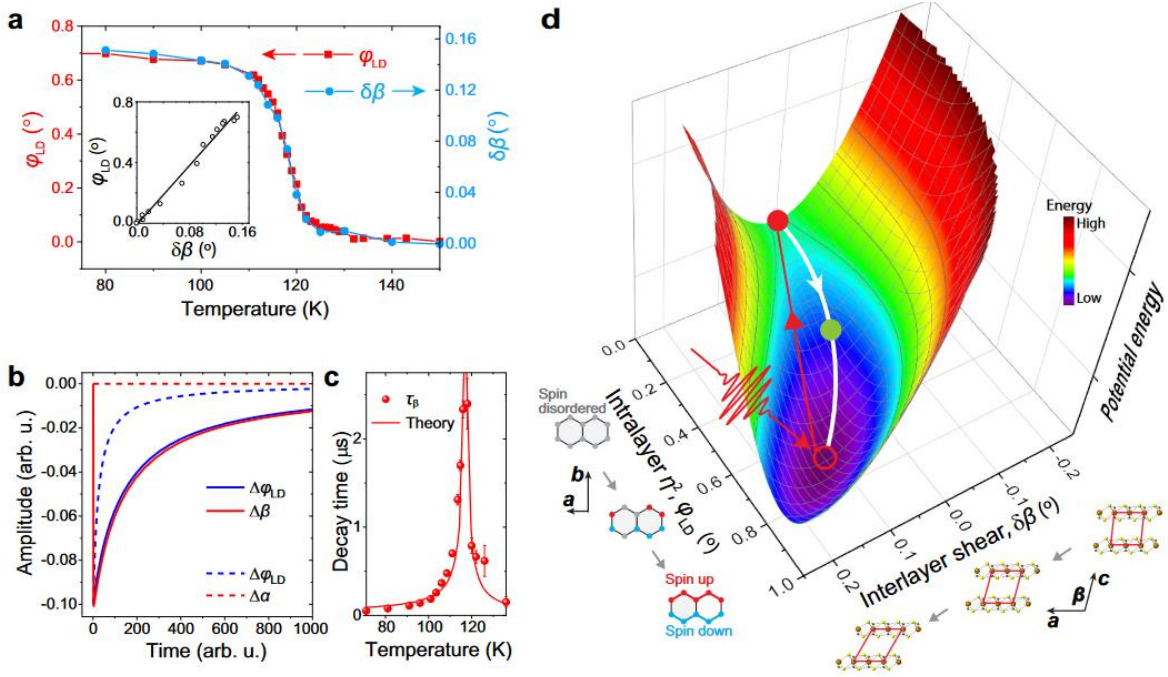

**Fig. 4 | The Ginzburg-Landau model with linearly coupled structural and magnetic degrees of freedom. a**, Static OLD polarization rotation $\varphi_{LD}$ and monoclinic angle change $\delta\beta$ in FePS$_3$ as a function of temperature. The inset shows the linear relation of these two parameters. **b**, Evolution of $\Delta\varphi_{LD}$ (i.e., $\eta^2(t) - \eta^2(t<0)$), $\Delta\beta$ (i.e., $\delta\beta - 0.1°$, $0.1°$ is the $\beta$ change across T$_N$.), and $\Delta\alpha$ close to T$_N$. Solid lines: coupled relaxation of $\Delta\varphi_{LD}$ and $\Delta\beta$ based on a model with a linearly coupled term ($\eta^2\delta\beta$). Dashed lines: decoupled relaxation of $\Delta\varphi_{LD}$ and $\Delta\alpha$ based on a model with a quadratically coupled term ($\eta^2\delta\alpha^2$, see Methods). **c**, The relaxation time of the interlayer shearing ($\tau_\beta$) agrees with the numerical results based on the Ginzburg-Landau model (solid curve) **d**, Schematic of the relaxation process of OLD polarization rotation and interlayer shear. The potential energy surface is generated from equation (1), shown as a function of intralayer magnetic order parameter and monoclinic angle. The system is excited from the ground state (open red circle) to the excited state (solid red dot) and relaxes to the ground state by a coupled recovery of magnetic order and monoclinic angle. The green solid dot represents the partially recovered intermediate state.

# Supplementary Information for

# Dynamical criticality of spin-shear coupling in van der Waals antiferromagnets

Faran Zhou[1], Kyle Hwangbo[2], Qi Zhang[1,2,9], Chong Wang[3], Lingnan Shen[2], Jiawei Zhang[1], Qianni Jiang[2], Alfred Zong[4], Yifan Su[5], Marc Zajac[1], Youngjun Ahn[1,6], Donald A. Walko[1], Richard D. Schaller[7], Jiun-Haw Chu[2], Nuh Gedik[5], Xiaodong Xu[2,3], Di Xiao[3,2], Haidan Wen[1,8*]

[1]X-ray Science Division, Argonne National Laboratory, Lemont, IL, USA

[2]Department of Physics, University of Washington, Seattle, WA, USA

[3]Department of Materials Science and Engineering, University of Washington, Seattle, WA, USA

[4]Department of Chemistry, University of California Berkeley, Berkeley, CA, USA

[5]Department of Physics, Massachusetts Institute of Technology, Cambridge, MA, USA

[6]Department of Materials Science and Engineering, University of Wisconsin-Madison, Madison, WI, USA

[7]Center for Nanoscale Materials, Argonne National Laboratory, Lemont, IL, USA

[8]Materials Science Division, Argonne National Laboratory, Lemont, IL, USA

[9]Present address: Department of Physics, Nanjing University, Nanjing, China.

*Correspondence to: wen@anl.gov

**Table of contents**

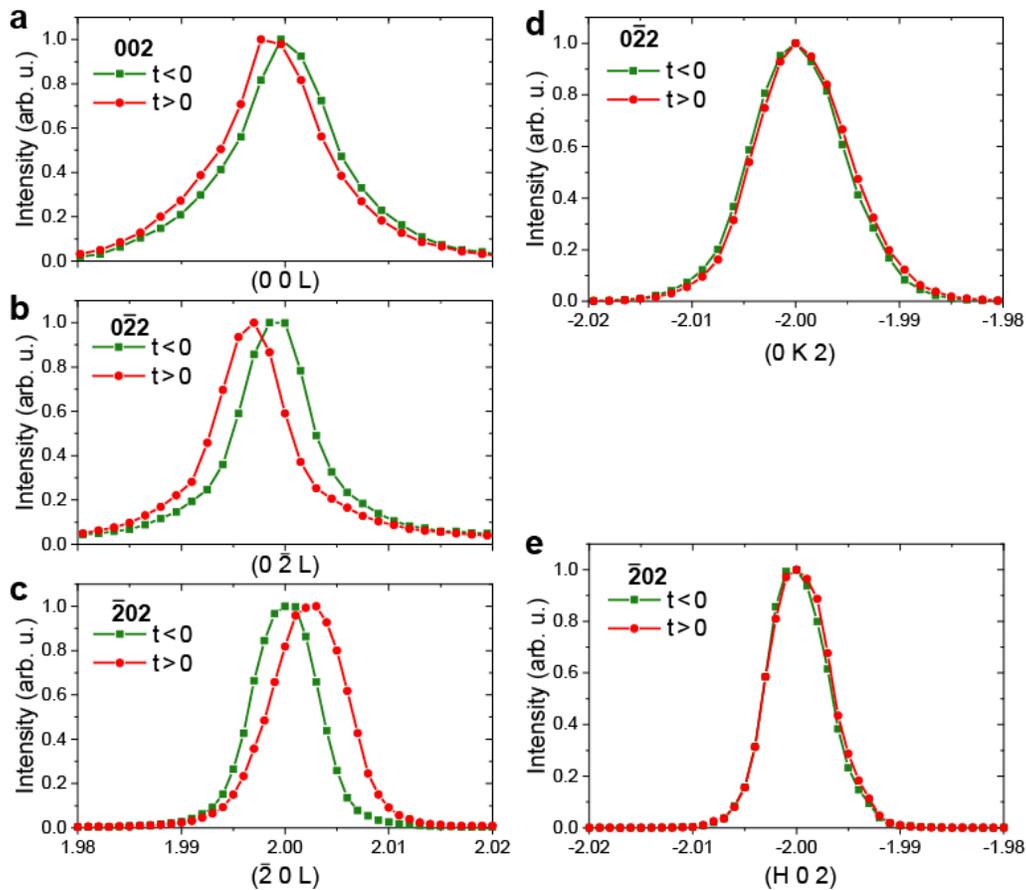

**Supplementary Figure 1**: HKL scans of 002, $0\bar{2}2$, $\bar{2}02$ Bragg peaks of FePS$_3$ before and after laser excitation. HKL scans of 002 (**a**), $0\bar{2}2$ (**b, d**), and $\bar{2}02$ (**c, e**) peaks before (t = -5 ns, green) and after (t = 25 ns, red) laser excitation at 95 K. The HKL scans were performed using SPEC program from Certified Scientific Software on a 6-circle diffractometer.

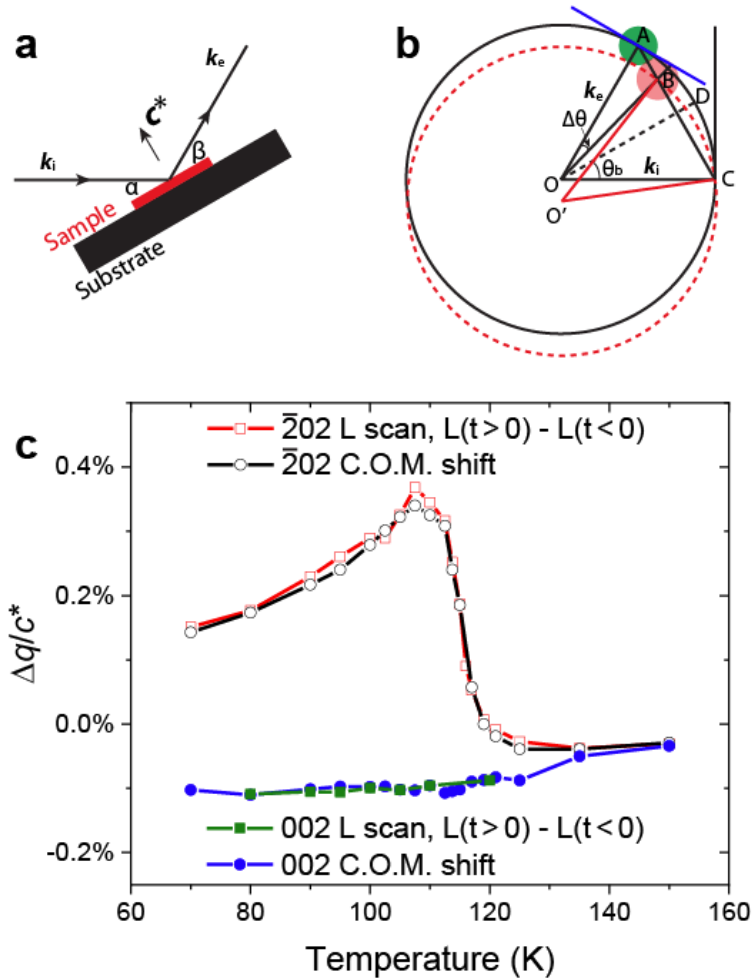

**Supplementary Figure 2**: Calibrating the peak shift measured by the center of mass on the detector in reciprocal unit. **a** Schematic of XRD reflection geometry. **b**, Schematic of the Ewald sphere and Bragg peaks in reciprocal space. Green and red disk: finite size Bragg peaks before and after laser excitation. Red dashed circle and red lines: Bragg condition for t > 0. **c**, The Bragg peak shifts as measured by the center of mass for 002 and $\bar{2}02$ peaks in FePS$_3$ are compared with the results obtained by the L scans. An example of L scan at 95 K is shown in Supplementary Figure 1.

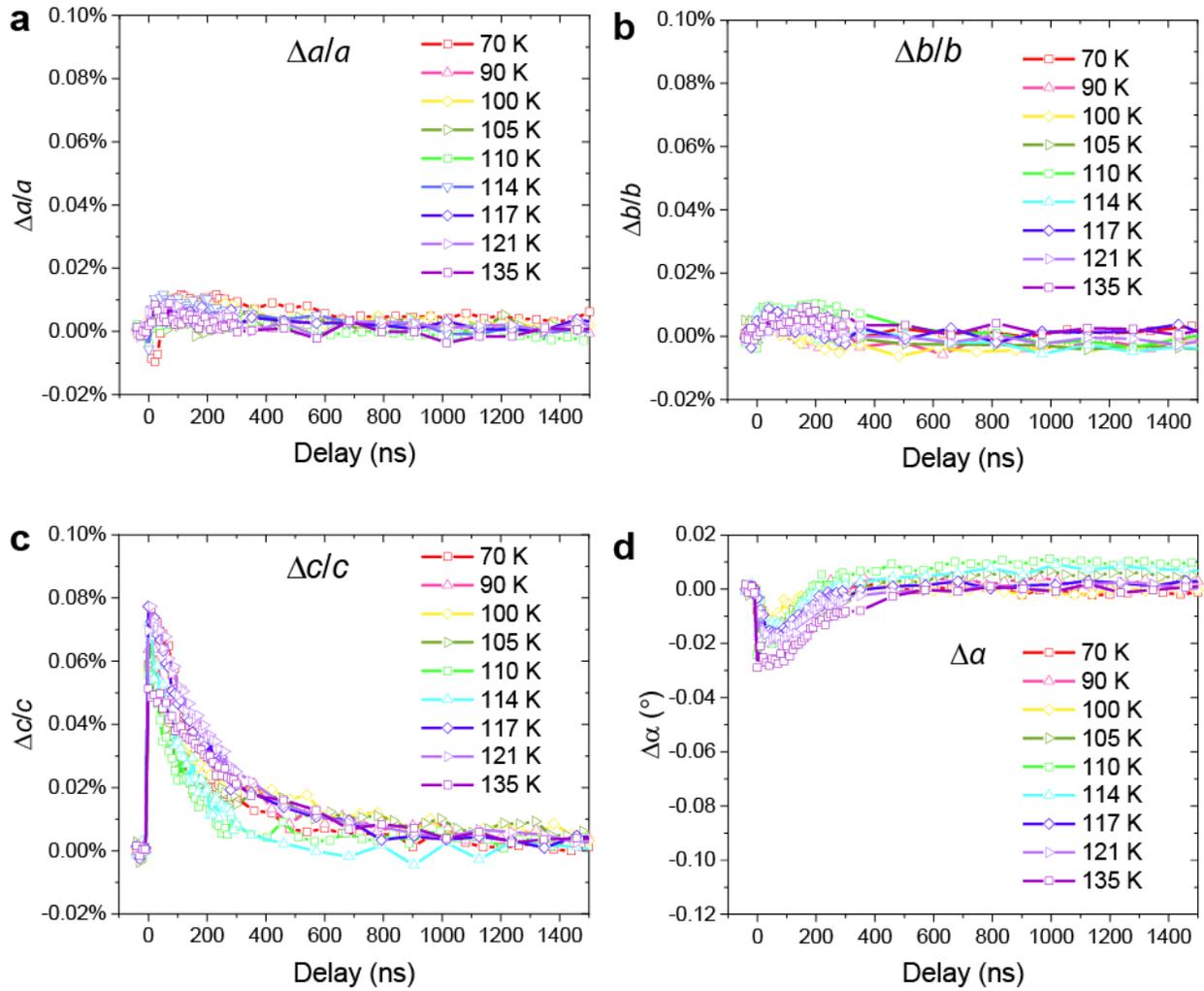

**Supplementary Figure 3**: Dynamics of FePS$_3$ lattice parameters $a$ (**a**), $b$ (**b**), $c$ (**c**), and $\Delta\alpha$ (**d**). These parameters were calculated based on the shifts of $\bar{2}02$, $0\bar{2}2$, and 002 Bragg peaks. Monoclinic angle $\beta$ was extracted from $\bar{2}02$ and 002 dynamics, detailed in Supplementary Note 3. Similarly, angle α was extracted from $0\bar{2}2$ and 002 dynamics. Lattice parameters were extracted based on: $a = \frac{2\pi}{\sin(\beta)\, a^*}$, $b = \frac{2\pi}{b^*}$, $c = \frac{2\pi}{\sin(\beta)\, c^*}$, $\Delta\alpha = \alpha(t) - \alpha(t < 0)$.

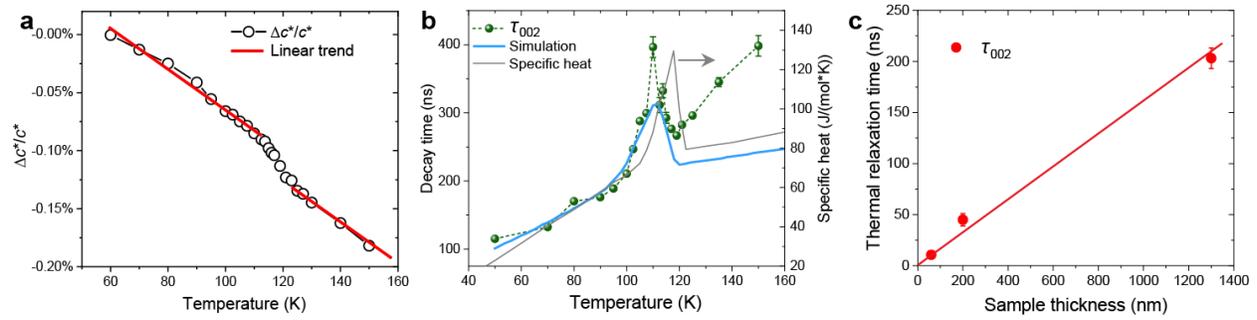

**Supplementary Figure 4**: Thermal expansion and relaxation of $FePS_3$. **a**, The measured percentage change of $c^*$ as a function of temperature. The red lines represent the linear thermal expansion at temperatures below and above $T_N$. **b**, Simulation of the thermal relaxation time. The gray line is the specific heat adapted from Ref.[1]. The green data points are the relaxation time measured by c* change of 002 peak. The blue curve is the simulation results based on a 1D thermal transport model as described in Supplementary Note 2. The error bars represent the standard deviation of exponential fitting. **c**, Thickness-dependent thermal relaxation time measured at 100 K. The red line is a linear fit of the data. The error bars represent the standard deviation of exponential fitting.

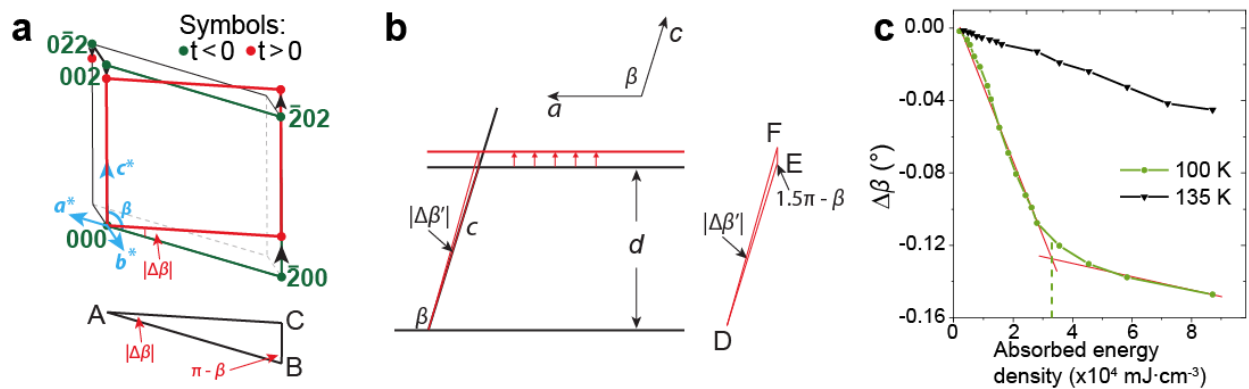

**Supplementary Figure 5**: Calculation of monoclinic angle change ($\Delta\beta$) and its dependence on absorbed energy density. **a**, Changes of $\beta$ based on 002 and $\bar{2}02$ peak shift. **b**, $\beta$ decrease due to an interlayer expansion (see Supplementary Note 3). **c**, Maximal $\Delta\beta$ at 100 K and 135 K as a function of absorbed energy density. The green dashed line indicates the threshold fluence at 100 K, as determined from the crossing of the linear trends (thin red lines) extrapolated from the low- and high-fluence limit.

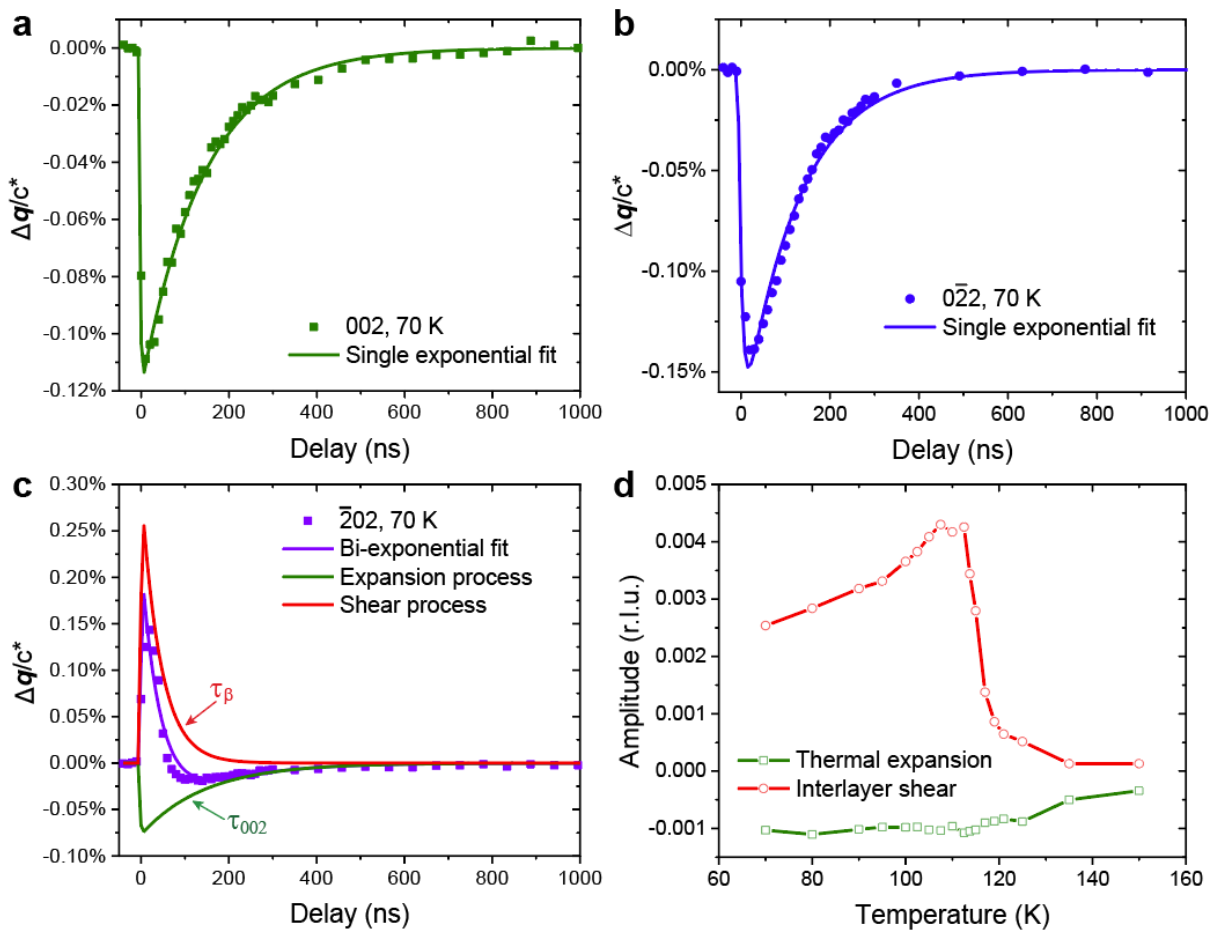

**Supplementary Figure 6**: Fitting of structural dynamics of FePS$_3$ by exponential decay functions. **a & b**, Single-exponential decay fitting of 002 and 0$\bar{2}$2 peak at 70 K, respectively. **c**, Bi-exponential decay fitting of $\bar{2}$02 peak at 70 K. The green curve corresponds to the relaxation of interlayer expansion measured by the 002 Bragg peak while red curve corresponds to the relaxation of the $\beta$ angle change. Purple curve is the sum of green and red curves. **d**, Comparison of the fitted amplitudes for thermal expansion (green curve) and interlayer shear (red curve) measured by 002 and $\bar{2}$02 peak as a function of the temperature.

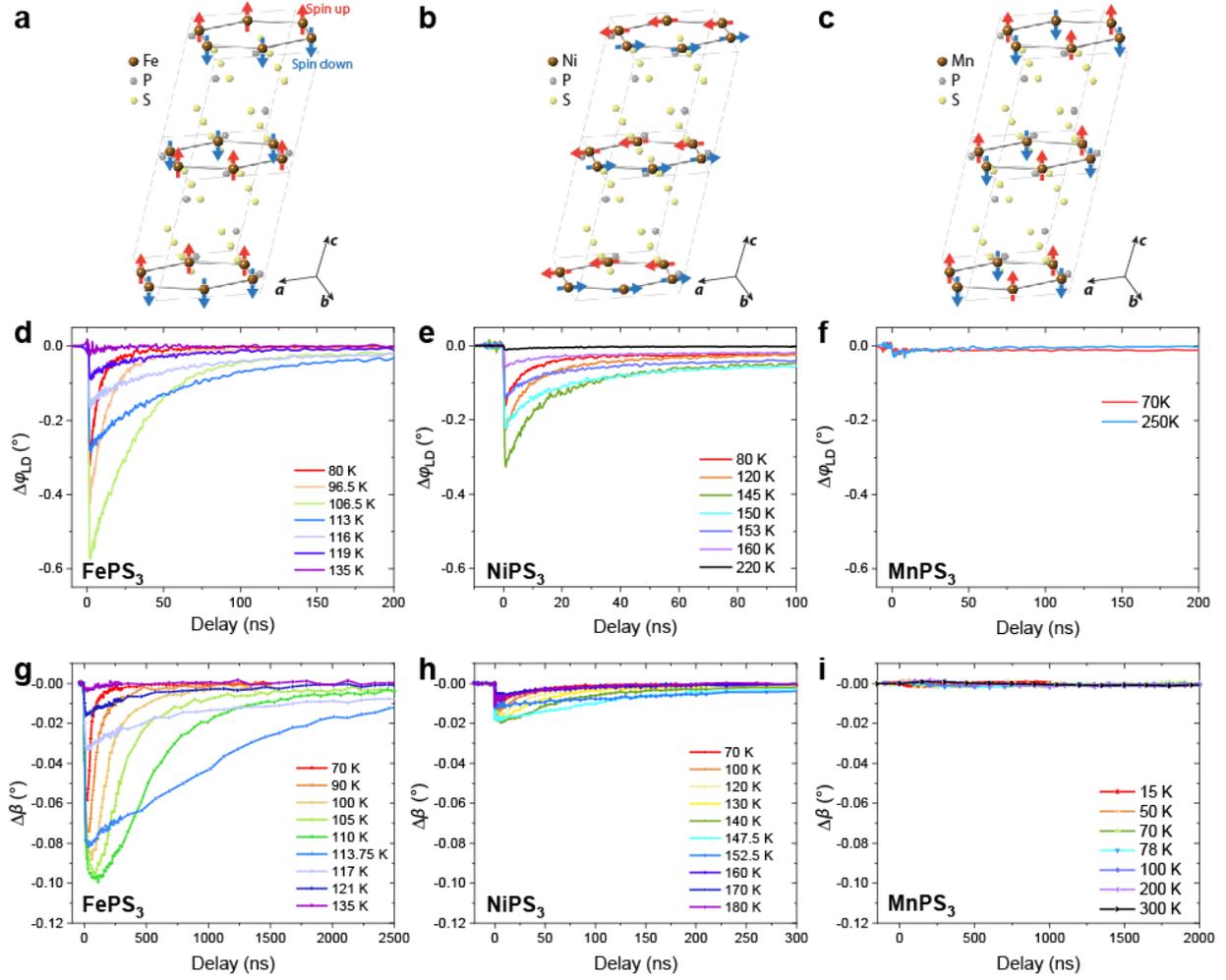

**Supplementary Figure 7**: Comparison of $\Delta\varphi_{LD}$ and $\Delta\beta$ dynamics in $M$PS$_3$ ($M$ = Fe, Ni, Mn). **a-c**, Crystal and magnetic structures of FePS$_3$, NiPS$_3$, and MnPS$_3$, respectively. **d-f**, $\Delta\varphi_{LD}$ dynamics in FePS$_3$, NiPS$_3$, and MnPS$_3$, respectively. **g-i**, $\Delta\beta$ dynamics in FePS$_3$, NiPS$_3$, and MnPS$_3$, respectively. The $\Delta\varphi_{LD}$ and $\Delta\beta$ dynamics are plotted in the same scale for the three compounds for comparison. The sample thicknesses are the following: FePS$_3$: 60 nm (**d**) and 1300 nm (**g**). NiPS$_3$: 40 nm (**e**) and 196 nm (**h**). MnPS$_3$: 50 nm (**f**) and 1100 nm (**i**). The sample thicknesses were chosen for optimizing the signal-to-noise ratios of the corresponding X-ray and optical probes. It is an extrinsic factor with a linear relation to the absolute delay time. But the thickness does not affect the scaling exponents as shown in Fig.3. The pump fluence for all the OLD measurements was ~1.0 mJ·cm$^{-2}$ (400 nm, 1 kHz repetition rate). The pump fluences for the XRD measurements was set such that the interlayer thermal expansion was similar ($\Delta c^*/c^* \approx 0.05\%$) for all three compounds.

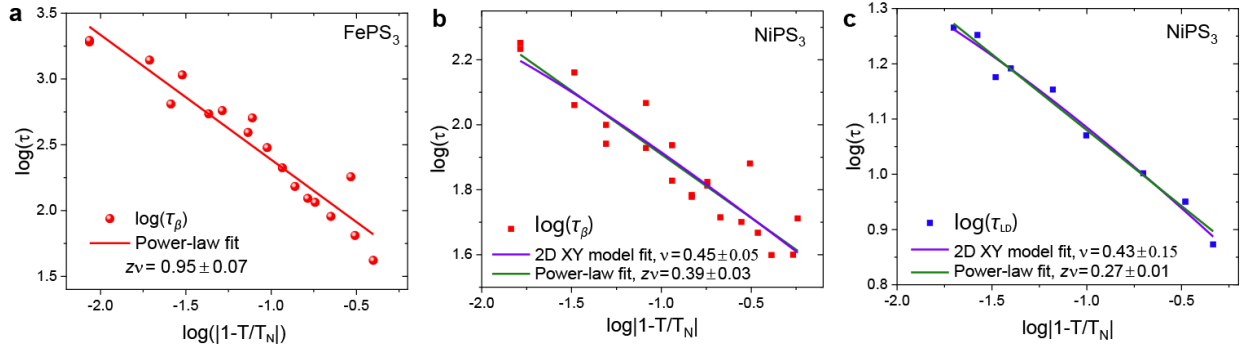

**Supplementary Figure 8**: Fitting of decay time as a function of temperature for FePS$_3$ and NiPS$_3$. **a**, Power-law fitting of $\tau_\beta$ for 1.3 μm-thick FePS$_3$. **b-c**, Power-law and 2D XY model fitting of $\tau_\beta$ (**b**) and $\tau_{LD}$ (**c**) for NiPS$_3$. Details of the fitting can be found in Supplementary Note 5. The quoted value of $z\nu=0.33$ for NiPS$_3$ in the main text is the average value shown in **b** and **c**. The quoted errors are fitting errors.

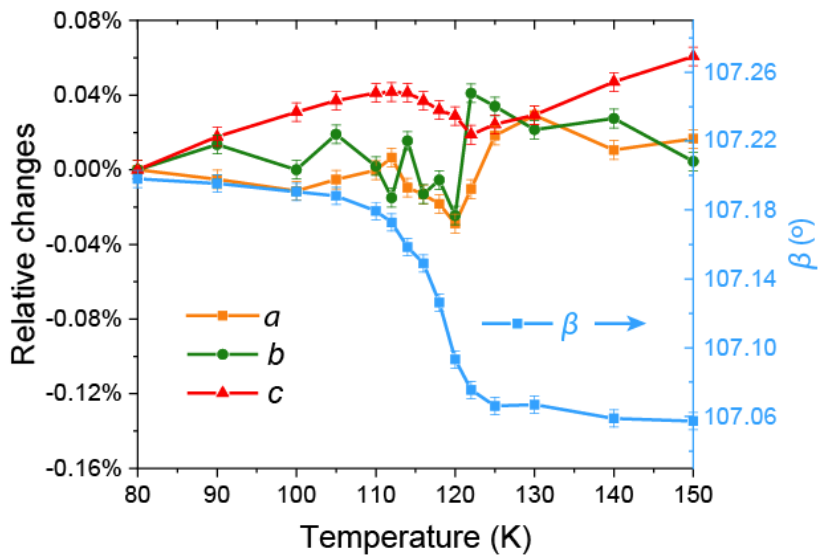

**Supplementary Figure 9**: Temperature dependence of the lattice parameters $a$, $b$, $c$, and $\beta$ based on the single-crystal XRD of $FePS_3$. The error bars indicate the uncertainties in determining the HKL values of the refined Bragg peaks.

**Supplementary Note 1: Comparison of peak center of mass shift and HKL scans**

The time-resolved XRD measurements were performed in such a scheme that the sample and the detector were fixed while the diffraction peak was recorded on an area detector as a function of time delays. We show below that the change of the center of mass on the detector at the fixed incident angle in our experimental setup can be used to measure the Bragg peak shift in reciprocal space.

As shown in Supplementary Figure 2a, before time zero, the Bragg condition is satisfied, i.e., $\alpha = \beta = \theta_b$, where $\alpha$, $\beta$, $\theta_b$ are the incident angle, exit angle, and Bragg angle, respectively. After laser excitation, the change of atomic plane spacing results in a change of the Bragg condition. As shown in Supplementary Figure 2b, the green and blue disks represent the finite-size Bragg peak. $\Delta\theta$ is the measured angular peak shift on the detector from point A to B. $\angle AOC = 2\theta_b$ is the $2\theta$ angle before laser excitation. $\angle BOC$ is the $2\theta$ angle after laser excitation. After laser excitation, $\angle BO'C$ is the theoretical $2\theta$ angle satisfying the Bragg condition, which can be calculated based on the peak shift on the detector. Based on trigonometry, one obtains

$$\angle BO'C = 2 * \sin^{-1}\left(\frac{BC}{2k}\right) = 2\sin^{-1}\left(\sin\theta_b - \frac{1}{2\sin\left(\frac{\pi}{2}+\theta_b-\Delta\theta\right)}\sin\Delta\theta\right)$$

In the experiment, $\theta_b$ for 002 peak is 10.15°. The measured peak shift on the detector at +5 ns corresponds to $|\Delta 2\theta| = 0.01°$. The theoretical angular shift $|\Delta 2\theta| = 2\theta_b - \angle BO'C = 0.0103°$, 3% larger than the measured peak shift of 0.01°. To confirm that the systematic error of measuring Bragg peak shift using diffraction peak center of mass change is negligible, the extracted shifts of 002 and $\bar{2}02$ peak were compared with the scans along the miller index L direction (L scan) using SPEC from Certified Scientific Software (Supplementary Figure 2c). As the figure shows, for both 002 and $\bar{2}02$ peaks, the peak center of mass shift is consistent with the L scan.

**Supplementary Note 2: Estimation of transient temperature rise and simulation of thermal relaxation**

The sample temperature rise upon laser excitation was estimated based on the interlayer lattice expansion. As shown in Supplementary Figure 4a, the measured change of $c^*$ (interlayer spacing $d = \frac{2\pi}{c^*}$) follows a linear dependence on the temperature above and below $T_N$. The kink at $T_N$ indicates a structural phase transition. The interlayer thermal expansion coefficient was estimated to be $1.818 \times 10^{-5}$ K$^{-1}$. In the pump-probe measurements, at a static temperature of 80 K, the measured Bragg peak shift $\Delta c^*/c^* = -0.055\%$ corresponds to a transient temperature increase of 30 K, which corresponds to an incident fluence of 5.7 mJ·cm$^{-2}$ on the 1.3 μm-thick sample.

A one-dimensional (1D) thermal transport model was employed to simulate the cooling of the sample including the latent heat during the phase transition, following the method detailed in Ref.[2]. Using the following parameters: sample interlayer thermal conductivity[3] $k = 0.85$ W · m$^{-1}$ · K$^{-1}$, sample-substrate interfacial thermal conductivity $G$=0.17 W · m$^{-1}$ · K$^{-1}$, the measured specific heat curve[1], the simulation results were summarized in Supplementary Figure 4b and captured the key features of experimental data. The thermal dissipation time gradually increases with sample temperature and peaks at 110 K, rather than $T_N$. The linear dependence of thermal relaxation time on sample thickness was also reproduced (Supplementary Figure 4c).

**Supplementary Note 3: Calculation of monoclinic angle $\beta$ change**

$\Delta\beta$ was calculated based on the 002 and $\bar{2}02$ peak shift. As shown in Supplementary Figure 5a, the 002 peak shift to the $-\mathbf{c}^*$ direction while the $\bar{2}02$ peak shift to $+\mathbf{c}^*$ direction. The difference between the two peak shifts is directly related to $\beta$ change. Based on the trigonometry shown in Supplementary Figure 5a, $|\Delta\beta| = \sin^{-1}(\frac{BC}{\sqrt{AB^2+BC^2-2*AB*BC*\cos(\pi-\beta)}}\sin\beta)$. In a small-angle approximation: $|\Delta\beta| \approx \frac{BC}{\sqrt{AB^2+BC^2+2*AB*BC*\cos\beta}}\sin\beta \propto BC$. The maximum BC we observed was 0.43%$c^*$ at 110 K, corresponding to $|\Delta\beta|=0.10°$ decrease as shown in Fig. 2d and Fig. 3c. The laser-induced $\beta$ decrease of 0.1° is consistent with the temperature-dependent XRD results, as shown in Supplementary Figure 9.

Since the crystal structure is monoclinic, an interlayer expansion with no interlayer shear can also lead to a decrease of $\beta$. Based on the trigonometry shown in Supplementary Figure 5b, $|\Delta\beta'| = \sin^{-1}(\frac{EF}{DF}\sin(3\pi/2 - \beta))$. Using a small-angle approximation, $|\Delta\beta'| \approx \frac{EF}{DF}\sin(3\pi/2 - \beta) \propto EF$. With EF = 0.05%$d$, we obtain $|\Delta\beta'| = 0.008°$, suggesting that when the interlayer spacing increases by 0.05%, $\beta$ decreases by 0.008° effectively, which is 12.5 times smaller than 0.1° $\beta$ change across the phase transition.

The $\beta$ change due to interlayer expansion was also observed experimentally. As shown in Supplementary Figure 5c, at the sample temperature of 135 K, $\Delta\beta$ increases linearly with the absorbed energy shown by the black curve. At the sample temperature of 100 K, the slope of $\beta$ change as a function of absorbed energy (the green curve) is 10 times larger than that measured at 135 K, corresponding to the ~10 times difference between total $\beta$ change and the interlayer expansion-induced $\beta$ change. Above the threshold of the absorbed energy density $3.3 \times 10^4$ mJ·cm$^{-3}$, the slope of the green curve matches the black curve measured as 135 K, suggesting that there is no longer interlayer shear and the $\beta$ change purely comes from interlayer expansion.

**Supplementary Note 4: Fitting of the time-dependent phenomena**

The relaxation time of XRD and OLD dynamics were extracted based on exponential function fitting. A single exponential decay function was used for fitting the time-dependent relaxation measured by 002 and $0\bar{2}2$ Bragg peak, as well as OLD, while a bi-exponential decay function was used to fit the dynamics measured by $\bar{2}02$ Bragg peak.

The single exponential decay function is: $\left(1 - e^{-\frac{t}{\tau_r}}\right)Ae^{-\frac{t}{\tau_d}}H(t)$, where $A$ is the amplitude, $t$ is the time delay, $\tau_r$ and $\tau_d$ are the rise and decay time constants, respectively. $H(t)$ is a step function that equals 0 when $t < 0$ and 1 when $t > 0$. To reduce the fitting parameters, the rise time of the dynamics, represented by $\tau_r$, was set to be a constant of 0.1 ns which matches the X-ray pulse duration. The amplitude $A$ and decay time constant $\tau_d$ were the fitting parameters. A representative fitting result is shown in Supplementary Figure 6a & b.

The bi-exponential decay function for fitting the dynamics of $\bar{2}02$ peak is: $\left(1 - e^{-\frac{t}{\tau_r}}\right)(A_1 e^{-\frac{t}{\tau_{002}}} + A_2 e^{-\frac{t}{\tau_\beta}})H(t)$. The two terms $A_1 e^{-\frac{t}{\tau_{002}}}$ and $A_2 e^{-\frac{t}{\tau_\beta}}$ describe the relaxation of the interlayer expansion and shear, respectively. The parameters of the first term were obtained from the fitting of the dynamics measured by 002 Bragg peak, while the amplitude and time constant of the second component were obtained by the least-squares fitting. At temperatures near $T_N$, the rise time of $\bar{2}02$ peak is longer than 0.1 ns so the rise time $\tau_r$ was also a fitting parameter, which does not affect the fitting results of the relaxation time constant A representative fitting result is shown in Supplementary Figure 6c. Below $T_N$, the interlayer shear has a much larger amplitude than thermal expansion (Supplementary Figure 6d) and dominates the relaxation process.

**Supplementary Note 5: Fitting of the critical exponents**

The divergence in of both $\tau_\beta$ and $\tau_{LD}$ as a function of temperature were fitted by a power-law function: $\tau = A|1 - \frac{T}{T_N}|^{-zv}$, where $A$ is the amplitude, $z$ and $v$ are the dynamical exponent and critical exponent of correlation length, respectively, T is the sample temperature. To better determine the sample temperature for each measurement, we first fit the data in a linear scale with three fitting parameters, $A$, $T_N$, and $zv$, which yield an initial value of $T_N$. We noted that this value has an offset of ±2 K with respect to $T_N$, which is attributed to the systematic error in the temperature measurements under laser illumination at different experimental conditions. Then, for better fitting of $zv$, the value of $T_N$ was fixed in the fitting procedure using the equation in the log scale: $\log \tau = -zv \log|1 - \frac{T}{T_N}| + \log A$, with only two fitting parameters: $zv$ and $A$. The fitted $zv$ values were summarized in Supplementary Table 1 for samples with various thicknesses, with a representative fit shown in Figure 8a.

The width of the temperature range in which the slowing down is observed can be determined by the temperature at which the critical fluctuation is larger than the thermal fluctuation. This region spans from 105 K to 117 K (Fig. 2e) as the recovery of lattice becomes much longer than the thermal recovery. The width of this region is wider than those observed such as in ultrathin Fe films[4] due to weak vdW interaction and large spin fluctuations in vdW magnets.

| Measurement | Sample thickness (nm) | $zv$ |
|---|---|---|
| XRD, $\beta$ | 60 | 0.95 ± 0.12 |
| XRD, $\beta$ | 200 | 1.15 ± 0.27 |
| XRD, $\beta$ | 1300 | 0.95 ± 0.07 |
| OLD, $\varphi_{LD}$ | 60 | 1.00 ± 0.04 |

**Supplementary Table 1**, Summary of critical exponents for different measurements in FePS$_3$. The error bar of $zv$ represents the standard deviation of the power-law fitting.

For NiPS$_3$, a 2D XXZ model system[5], the relaxation time as a function of temperature can take the following form[6]: $\tau = a|1 - \frac{T}{T_N}|^{-v} e^{bz|1-\frac{T}{T_N}|^{-v}}$, where $a$, $b$ are constants, $z$ is the dynamical exponent, $v$ is the critical exponent of the correlation length. For 2D XY model fitting, only $v$ rather than $zv$ was obtained from the fitting (Supplementary Figure 8b & c). The fitted value was $v = 0.44 ± 0.11$, close to the theoretical value of $v = 0.5$ (Ref.[6]).

As shown in Supplementary Figure 7, the shear amplitude and the magnetic order measured by OLD are larger in FePS$_3$ than that in NiPS$_3$. A possible explanation is the different electron configurations of the corresponding ions. For example, the Fe$^{2+}$ ion has d$^6$ configuration, which leaves the minority $t_{2g}$ manifold partially filled, but the $Ni^{2+}$ ion has $d^8$ configuration, so the minority $t_{2g}$ manifold is fully filled, which quenches the angular momentum and in general leads to weaker spin-lattice coupling compared to partially filled d-orbitals. In addition, the strength of critical slowing down is also different: the recovery time of the shear motion in $FePS_3$ increased by two orders of magnitude and only increased by one order of magnitude in $NiPS_3$. This is likely due to the different spin configurations in these two compounds. The zigzag spin order is Ising type and 2D XY type in $FePS_3$ and $NiPS_3$ respectively, which have distinct scaling exponents and universality classes.

**Supplementary Note 6: Temperature-dependent single-crystal XRD for FePS$_3$**

We performed single-crystal XRD as a function of temperature at the 7ID-C beamline of the Advanced Photon Source. During the measurement, the orientation matrix was optimized and fixed at 80 K. The H, K, L values of multiple Bragg peaks (002, 004, $\bar{2}$04, $\bar{4}$24) were recorded at each temperature and used to calculate the *a*, *b*, *c*, and *β*, as summarized in Supplementary Figure 9.

We note that previous reports of the structural transition in FePS$_3$ used the powder XRD method[7,8] did not report monoclinic angle *β* change as a function of temperature. To cross check our single-crystal structure refinement results, we also performed high-resolution powder XRD at 11BM beamline at the Advanced Photon Source. The powder XRD results yield a distinct *β* change of 0.02° across $T_N$. This is a factor of 5 smaller than the single-crystal result. We ascribe this discrepancy to the sample preparation for powder diffraction, in which the samples were grinded into powder and attached to the Kapton tubes with vacuum grease. This process likely introduced strains in the sample, resulting in a smaller *β* change in powder XRD data.

**Supplementary References**